\documentclass[a4paper,12pt,final]{article}
\usepackage{latexsym,amssymb,epsfig,graphicx}

\begin{document}

\title
{Physics of Cosmic Reionization}
\author{T. Roy Choudhury$^{1,2}$\thanks{E-mail: chou@sissa.it}~
and
A. Ferrara$^{1}$\thanks{E-mail: ferrara@sissa.it}\\
\footnotesize{$^{1}$SISSA/ISAS, via Beirut 2-4, 34014 Trieste, Italy}\\
\footnotesize{$^{2}$Centre for Theoretical Studies, Indian Institute of Technology,
Kharagpur 721302, India}
}
\date{}

\maketitle

\begin{abstract}
The study of cosmic reionization has acquired increasing significance over the last few years
because of various reasons.
On the observational front, we now have good quality data of different types 
at high redshifts (quasar absorption spectra, radiation backgrounds at different frequencies, cosmic microwave background polarization, Ly$\alpha$ emitters and so on). Theoretically, the importance of the reionization
lies in its close coupling with the formation of first cosmic structures, and there have been 
numerous
progresses in modelling the process. In this article, we review the 
current status of our understanding of the physical processes governing the cosmic
reionization based on available observational data.

\end{abstract}


\section{Historical background}
\label{intro}

Reionization can be thought of as the second major change in the ionization
state of hydrogen (and helium) in the universe (the first being the 
recombination occurring at $z \approx 1100$). The process is 
of immense importance in the study of structure formation since, on 
one hand,  
it is a direct consequence of the 
formation of first structures and luminous sources while, on the other, 
it affects subsequent structure formation.
In this article, 
we attempt to review the basic physical processes related to the
reionization with particular emphasis on the link between theory and observations. 

The study of reionization consists of two broad areas, 
namely the properties of the intergalactic medium (IGM) and the 
formation of sources. 
Once the first sources produce photons capable of ionizing the
surrounding IGM, the process of reionization (and reheating) 
can be thought of having begun, thus changing the thermal, ionization 
and chemical properties of the IGM.
This change in the nature of the IGM affects the 
formation of next generation of sources (like 
metal enrichment changing the initial mass function of 
stars).
The subject area of formation of sources is quite involved in itself
dealing with formation of non-linear structures (haloes and filaments), 
gas cooling, accretion and generation of radiation (stars and quasars).
This is somewhat beyond the scope of this review, and 
we will mostly concentrate on the effects of sources on the IGM. 

Historically the study of reionization of the IGM has been
closely linked 
with the observations related to spectra of distant 
quasars, in particular to the Ly$\alpha$ forest, though it was not 
obvious in the beginning whether the Ly$\alpha$ forest traces baryons
of cosmological significance.
In particular, models in which the Ly$\alpha$ forest arises
from some kind of ``confined clouds'' predicted that 
the amount of baryons within the 
forest may not be of cosmological significance and hence
may not have any substantial connection to the cosmic reionization
as we understand it today. 
To stress this in slightly more detail, let us briefly review some of the 
major ideas in the development of this field. Of course, the literature
of the Ly$\alpha$ forest has been reviewed various times, 
(see for example \cite{rauch98}); nevertheless it might be appropriate
to review some the major ideas from the point of view
of cosmic reionization.

\subsection{Initial models based on pressure and gravitational confinement}

In a classic paper, Gunn \& Peterson \cite{gp65} showed
that the hydrogen in a diffuse uniform IGM must have 
been highly ionized at $z \approx 2$ in order to 
avoid complete absorption of the transmitted flux 
at wavelengths bluewards of the Ly$\alpha$ emission line
of the QSO; this is now commonly known as the 
Gunn-Peterson (GP) effect. Following that, it was proposed
\cite{bs66} 
that this GP effect can be used to probe the ionization state
of hydrogen within the IGM at various redshifts 
(and also for other elements). The GP effect has remained one of the 
most stringent tests of the ionization state of the IGM to date.

At the same time, it was also realized that gas which
was not uniformly distributed would produce discrete Ly$\alpha$ absorption 
lines. In the beginning, 
the most natural structures considered were gas clumped 
into groups of galaxies \cite{bs65} 
or low mass protogalaxies \cite{arons72}. However, these
were soon found to be unrealistic when different groups
\cite{lynds71,sybt80}
discovered a large number of discrete absorption lines in the QSO spectra,
which are usually known as the ``Ly$\alpha$ forest''.
It was shown that these
forest lines could not be associated with galaxy clusters, rather 
they have an intergalactic origin and arise in
discrete intergalactic clouds at various cosmological 
redshifts along the line of sight (for reviews 
see \cite{sybt80,wcs81,btn88}).  
Various arguments 
(like the apparent lack of rapid evolution in the properties of the forest, 
the short relaxation time scales for electrons and 
protons 
and short mean free paths) led to the notion that the clouds 
were ``self-contained
entities in equilibrium'' \cite{sybt80}. 
A two-phase medium was postulated,
with the diffuse, very hot, intercloud medium (ICM) in 
pressure equilibrium with the
cooler and denser Ly$\alpha$ clouds. 
In this two-phase scenario, the ICM was identified with the IGM, while
the Ly$\alpha$ clouds were treated as separate entities.

According to the pressure confinement
model \cite{sybt80,oi83,io86}, the Ly$\alpha$ 
clouds are supposed to be in photoionization equilibrium with an ionizing 
ultra-violet (UV) background. The gas is heated by
photoionization and cools via thermal 
bremsstrahlung, Compton cooling, and
the usual recombination and collisional excitation processes.  
Since the ICM is highly ionized, the photoheating is not efficient and hence
the medium cools adiabatically through cosmic expansion. 
The denser clouds embedded in the hot ICM
have a nearly constant temperature fixed by 
thermal ionization
equilibrium ($\sim 3 \times 10^4$ K) \cite{oi83,io86}.
The available range of cloud masses
is constrained by the requirement that the clouds must be small enough not to
be Jeans-unstable but large enough not to be evaporated rapidly when heated
by thermal conduction from the ambient ICM \cite{sybt80,oi83}. 
According to such constraints, 
clouds formed at high redshifts 
would survive down to observed
redshifts only if their masses range between $10^5$--$10^{10} M_{\odot}$.

The neutral hydrogen within the confining ICM is expected to cause 
a residual GP absorption trough between the absorption
lines (clouds). 
However, observations at higher spectral resolution 
\cite{ss87,gct92,gdf++94}
revealed no continuous absorption between the discrete
lines, placing strong limits on the GP effect, which in turn, puts a
strict
upper limit on the density 
of the ICM. The ICM temperature
has a lower limit from the absorption line width,
while
the condition that the cloud must be large enough not to evaporate gives an
upper limit on the temperature \cite{oi83}. 
Another independent upper limit on the temperature
of the ICM comes from the lack of inverse Compton distortions in the spectrum
of the cosmic microwave background \cite{bfr91}
through the
Sunyaev-Zeldovich effect \cite{sz80}. 
In fact, the upper limit of the so-called 
$y$-parameter \cite{fcg++96}
is able to rule out any cosmologically distributed component 
of temperature greater than $10^6$ K.
When all the limits are combined, only a
relatively small corner of allowed density-temperature parameter 
space remains for the ICM. It turns out that, according to the
pressure-confinement model, the density of the ICM is too small
to be cosmologically significant. Hence, during these early days,
the connection between the cosmic reionization and the IGM was
not at all obvious as most of the baryons was expected to lie
somewhere else.

The pressure-confinement model ran into severe
problems while trying to match the observed 
column density distribution \cite{pwrcl93}. 
For example, in order 
to reproduce the low column density systems between, say, 
$13 < \log(N_{\rm HI}/{\rm cm}^{-2}) < 16$
(where $N_{\rm HI}$ is the column density of neutral hydrogen), 
the mass has to vary by 9 orders of magnitude.  On the other hand,
the mass is severely constrained in order to ensure cloud survival.
Therefore, the only escape route is to invoke pressure inhomogeneities 
\cite{bchw89}.
However, the Ly$\alpha$ absorbers are found to be weakly clustered
over a large range of scales, which thus excludes any significant
pressure fluctuations \cite{wb91}. 
Similarly, detailed hydrodynamical simulations \cite{wb92}
show  that the small mass range of the clouds leads to a failure in producing
the column density distribution at high $N_{\rm HI}$. 
In addition, pressure-confinement models predict small cloud sizes which are 
incompatible with the observations of multiple lines of sight 
\cite{dfi++95}.
It was thus concluded that 
the pure pressure confinement model is unlikely to explain
the Ly$\alpha$ forest as a whole though it is possible that some 
lines of sight must go through
sites where gas is locally confined by external pressure
(say, the galactic haloes, the likely hosts of the dense Lyman limit
absorbing clouds). 

Even from a theoretical point of view, there are no physical reasons for
preferring pressure to gravitational confinement or to no confinement at all. 
Because of this, 
self-gravitating baryonic clouds were suggested by 
\cite{melott80,black81} as an 
alternative to the pressure confinement model. 
In this model, the appearance of
the IGM as a forest of lines is because of the variations in the
neutral hydrogen density rather than a sharp transition between 
separate entities. In this sense, there is no real difference
between an ICM and the clouds in the gravitational confinement model. 
This scenario of self-gravitating clouds 
predicts larger sizes of the absorbing clouds
($\sim 1$ Mpc) compared to the pressure-confinement scenario.
However, this model, too, runs into problems while trying
to match the observed column density distribution \cite{pbcp93} as 
it predicts larger number of high
column density systems than is observed. Secondly, the large
absorber sizes seemed to contradict observations. Furthermore, 
gravitationally confined clouds are difficult to explain theoretically since 
the mass of such clouds must lie in a restricted range to maintain 
the gas in equilibrium against free expansion or collapse.

As a further alternative, 
the properties of gas clouds confined by the
gravitational field of dark
matter have been investigated \cite{ui85}, more 
specifically in terms of the ``minihalo'' model \cite{rees86,ikeuchi86}.
In this picture, Ly$\alpha$ clouds are a natural byproduct 
of the cold dark matter (CDM) structure formation
scenario. Photoionized gas settles in the potential well of an isothermal 
dark matter
halo. The gas is stably confined if the potential is sufficiently 
shallow to avoid
gravitational collapse but deep enough to prevent the warm gas from escaping.
CDM minihaloes are more compact than the self-gravitating baryonic clouds of
\cite{black81} because of the larger dark matter gravity, thus alleviating the 
size problem. The detailed structure
of the halo depends on the relative spatial distribution of baryons and CDM.
However, the virial radii of the confining objects ($\sim 10$ kpc)
are much lower than the coherence lengths of the Ly$\alpha$ systems
as obtained from constraints on absorption
line observations of lensed or paired QSOs \cite{sss++92,srs++95}.
It was thus natural to extend the minihalo model to non-static systems.
A non-static minihalo model was studied by \cite{bss88}, 
who examined the hydrodynamics of a 
collapsing spherical top-hat perturbation and suggested that clouds were in a
free expansion phase. 

\subsection{IGM as a fluctuating density field}

Following the non-static models, it was realized that 
an IGM with the density
fluctuation variance of the order of unity
could also produce line-like absorptions in quasar spectra
\cite{mcgill90,bbc92}.
According to such models, 
the IGM becomes clumpy and acquires peculiar motions under the
influence of gravity,
and so the Ly$\alpha$ (or GP) optical depth should vary 
even at the lowest column
densities \cite{black81,mcgill90,bbc92,mr93,rm95}. In a CDM-dominated structure
formation scenario, the accumulation of matter in overdense regions reduces
the optical depth for Ly$\alpha$ 
absorption considerably below the average in most
of the volume of the universe, leading to what has been called the fluctuating
GP phenomenon. Traditional searches for the GP effect that try to measure the
amount of matter between the absorption lines were no longer meaningful, as
they were merely detecting absorption from matter left over in the 
most underdense
regions. If this is not taken into account, the amount of ionizing radiation
necessary to keep the neutral hydrogen GP absorption below the 
detection limits can be overestimated, which would then
have severe implications for reionization studies.
In this scenario, the density, temperature and thermal pressure of the
medium were described as continuous fields and could not be attributed
simply to gravitational confinement or pressure confinement. 
These studies led to a shift
in the paradigm of IGM theories, 
especially since they implied that the IGM contains most of the baryons
at high redshifts, thus
making
it cosmologically significant and hence quite relevant to cosmic
reionization. 

The actual fluctuation picture can be derived from cosmological 
$N$-body and hydrodynamical simulations. 
It was possible to solve hydrodynamical equations
from first principles and set up an evolutionary
picture of the IGM in these simulations 
\cite{cmor94,zan95,hkwm96,mcor96}. 
Although different techniques
and cosmological models were used by different groups, 
all the simulations indicate a
fluctuating IGM instead of discrete clouds. 

Since in this new paradigm, the Ly$\alpha$ forest arises from a 
median-fluctuated quasi-linear
IGM, it is possible to ignore the high non-linearities.
This made it possible to study the IGM through semi-analytical
techniques too 
\cite{bbc92,bi93,gh96,bd97,hgz97}.
The issue of dealing with quasi-linear
densities were dealt in two ways. 
In the first method, it was showed that a quasi-linear 
density field, described by a lognormal
distribution, can reproduce almost all the observed 
properties of the Ly$\alpha$ forest
\cite{bbc92,bd97}. In fact, this was motivated by
earlier ideas of \cite{cj91} for dark matter distribution.
In an alternate method, it was also 
possible to obtain the 
density distribution of baryons from simulations which could
then be used for semi-analytical calculations \cite{mhr00}.
Given the baryonic distribution, the neutral hydrogen fraction was
calculated assuming photoionization
equilibrium between the baryons and the ionizing radiation
field.
It was also
realized that the equilibrium 
between photoheating and adiabatic cooling
implies 
a tight relation between the temperature and density
of the gas, described by a power-law equation of state 
\cite{gh98}, which was used for determining the
temperature of the gas. Given such simplifying 
and reasonable assumptions, it was possible to make
detailed predictions about the Ly$\alpha$ forest.
For example, a relation between column density peaks (``absorption lines'') 
and the statistics
of density peaks was proposed \cite{gh96,hgz97}, 
and analytical expressions for the dependence of
the shape of the column density distribution on 
cosmological parameters were obtained.
                                              
The simulations and the semi-analytical calculations
both have been quite successful
in matching the overall observed properties of the absorption systems. 
The shape of the 
column density distribution and the Doppler parameter distribution are reasonably
well reproduced by the simulations 
\cite{cmor94,hkwm96,mcor96,zan95,mpkr96,zanm97}
as well as semi-analytical calculations \cite{hgz97,cps01} over
a wide redshift range.
The large
transverse sizes of the absorbers seen against background paired and lensed
QSOs are well explained by the coherence length of the sheets and filaments
\cite{mcor96,cs97,cazn97}. 
In addition, the probability distribution function and 
power spectrum of the transmitted flux in the Ly$\alpha$ 
forest is reproduced very well by
the models \cite{mmr++00,csp01}. 
The Ly$\alpha$ optical depth fluctuations 
were used for recovering the power spectrum of matter density 
fluctuations at small
scales \cite{cwkh98,cwphk99} 
and also to  obtain various quantities related to the IGM
\cite{csp01,wmhk97}.  

Given the fact that the Ly$\alpha$ can be modelled so accurately, it has become
the most useful tool in studying the thermal and ionization history
of the universe ever since. 
Subsequently it was realized that this simple description of the IGM
could be coupled to the properties of the ionizing sources and 
hence it was possible to compute the reheating and reionization
history. Since the modelling of the sources is a highly non-linear
problem and much more non-trivial to solve that the 
quasi-linear IGM, it was more natural to make
some simple assumptions about the sources, calculate
their effect on the IGM and then constrain the properties
of the sources themselves.

\subsection{Sources of ionization}

The classic problem of the propagation of ionization fronts from 
a point source was studied by \cite{sg87,gs96}. 
It was shown that the recombination timescale is too large 
for the ionized region to reach the Str\"omgren radius. Furthermore, 
the calculations showed 
that the ionizing photons from the observed population of 
QSOs cannot produce enough UV flux to
reionize the IGM at $z \approx 3$ \cite{sg87,ds87,gs96}. 
This lead for extensive 
searches and proposals for other sources of UV ionizing flux.
The next most obvious choices for UV radiation
were the (early) galaxies and stars.
This was studied 
using observed ionization state of heavy element absorption 
systems in the spectra of QSOs and model-dependent metal production arguments 
\cite{ss89,scl90}, though
no firm conclusions could be drawn 
because of the fraction of photons which are able 
to escape the host galaxy is unknown
(and that situation remains till date).

The possibility of galaxies contributing to the UV flux was implemented
in various analytical calculations
\cite{mo90,madau91,bvn92,og96}. These 
calculations concentrated on the collapse of dark matter haloes,
subsequent cooling (atomic and/or molecular) of gas, star formation
formalisms and propagation of ionization fronts. 
Subsequently, detailed modelling for the reheating and
reionization histories of the IGM showed that, under standard
assumptions regarding hierarchical CDM model, 
Press-Schechter theory, cooling within 
collapsed haloes, star-forming efficiency and observed QSO luminosity
function, the reionization of the hydrogen is achieved
at $z \approx 10$ \cite{vs99,co00}. Most of these studies
generally incorporated the inhomogeneities in the IGM through
a (evolving) clumping factor. A model for reionization
for the inhomogeneous IGM was proposed \cite{mhr00} which
was able to take into account the fact that dense
regions would remain neutral longer (because of their high recombination
rate).

Most of these effects were also seen in hydrodynamical simulations, thus
confirming the overall picture for reionization by UV sources. 
Usually the limitation in computing power forced 
 small volumes 
(say, boxes with sizes of a few Mpcs) to be simulated. It was found 
\cite{og96,go97} that
a mass
 resolution of about $10^4 M_{\odot}$ was required to resolve early
 epochs of reheating and reionization, which remains a great challenge even
now. A better resolution can be achieved if, for example, 
high-resolution $N$-body simulations and semi-analytical models
for galaxy and star formation are combined \cite{cfgj00}
to obtain the 
thermal history of the IGM.

The picture of reionization by UV sources 
which emerged form these studies can be summarised
as follows: (i) The reionization process by UV sources could be classified
into three phases \cite{gnedin00}. In the ``pre-overlap'' phase, the ionized
regions of individual sources propagate into the neutral IGM. 
In the ``overlap'' phase, the ionized regions start overlapping and 
subsequently ionize the whole of IGM (except for some high-density
peaks). At this stage the universe becomes transparent to UV
radiation and hence the mean free path of photons increases
dramatically. Finally, there is the ever-continuing 
``post-overlap'' phase where the ionization
fronts propagate into the neutral high density regions. 
(ii) The reheating of the IGM preceded the reionization
as a small number of hard photons could heat the medium up to several
hundred to thousand Kelvins before complete reionization,

It should be mentioned that though the QSOs and galaxies seem to be
the most natural choices as sources for reionization of the IGM, 
the possibility of other sources cannot be ruled out, at least 
from observations. Hence various other sources
have been studied too, the early ones being 
the supernova-driven winds \cite{tse93}, 
hard photons from structure formation \cite{st94} 
early formed massive black holes \cite{su96}
and
more exotic sources like decaying dark matter (or other) 
particles \cite{sciama88,dj92,dj94,sciama94,sciama95,sn97}, 
with the list ever-increasing till date \cite{hh04,mf05}.

It is thus clear that the transmission regions in the 
Ly$\alpha$ forest at redshifts $z \lesssim 6$
conclusively implies that the universe is ionized at lower redshifts, 
though the exact nature of the ionization process or the sources responsible
are not understood at the moment. On the other hand, we can also
think of the Ly$\alpha$ forest as the leftover of the reionization, i.e.,
the absorption signatures imply 
that the sources were not able to fully complete the job.

In the next section, we shall review the current observational
situation regarding reionization of the IGM and main conclusions
that can be drawn from the data.
Section 3 would discuss the physics of cosmic reionization 
along with description of certain analytical and numerical models.
We shall summarize the main predictions and future tests
for these models.

\section{Observational constraints}

In this Section, we summarise various sets of observational data which 
shape our current understanding of reionization
(for a detailed review on recent developments, 
see \cite{fck06}). These 
observational probes can be broadly divided into two types: the 
first set probes the extent and nature of the reionization 
through observations of the IGM while the second is mostly
concerned with direct observations of the sources responsible
for reionization. 

\subsection{Observations related to the state of the IGM}
 
As far as the IGM is concerned, the observational constraints
on its ionization and thermal state can be divided broadly into
three classes, which are discussed in the next three subsections. 

\subsubsection{QSO absorption lines}

We have discussed in the previous Section that the primary evidence
for the IGM to be ionized at $z < 6$ comes from the 
measurements of GP optical depth in the spectra of QSOs.
Under the assumptions of photoionization equilibrium and 
a power-law relation between temperature and density, 
the Ly$\alpha$ optical depth $\tau_{\rm GP}$ arising from a region 
of overdensity $\Delta$ at a redshift $z$ can be written as
\begin{equation}
\tau_{\rm GP} = 3.6 \times 10^5 \left(\frac{\Omega_b h^2}{0.022}\right)
\sqrt{\frac{0.15}{\Omega_m h^2}} \left(\frac{1-Y}{0.76}\right)
\left(\frac{1+z}{7}\right)^{3/2} \bar{x}_{\rm HI} ~ \Delta^{\beta}
\label{eq:tau_gp}
\end{equation}
where $Y$ denotes the helium mass fraction, and  $\bar{x}_{\rm HI}$
is the neutral hydrogen fraction (defined as the ratio between
neutral hydrogen density and total hydrogen density) at the mean
density $\Delta = 1$. The exponent of $\Delta$ is 
determined by the photoionization equilibrium and is given by 
$\beta = 2.7 - 0.7 \gamma$, where $\gamma$ is the slope of the 
pressure-density relation. For a isothermal medium $\gamma=1$ and
hence $\tau_{\rm GP} \propto \Delta^2$.
All other symbols in the above expression have standard meanings.
This expression clearly shows that for a uniform
medium ($\Delta  = 1$) at $z \lesssim 6$, the presence of 
a neutral hydrogen fraction $\bar{x}_{\rm HI} \gtrsim 10^{-5}$
would produce an optical depth of the order unity and hence would show
clear GP absorption trough in the spectra. Since such absorption is not
observed for QSOs at $z < 6$, 
the constraint on the average neutral fraction is 
$\bar{x}_{\rm HI} < 10^{-5}$, which is a robust indication of the fact
that the universe is highly ionized at $z < 6$.

The observational situation changes for the observed QSOs at $z > 6$.
The ongoing Sloan Digital Sky Survey (SDSS)\footnote{http://www.sdss.org/}
has discovered quite
a few QSOs at $z \gtrsim 6$, the spectra of which are markedly
different from their low-redshift counterparts.
Very long absorption 
troughs, which are of the size $\sim$ 80--100 comoving Mpc, 
have been seen along tens of lines of sight at $z > 6$
\cite{bfw++01,fnl+01,dcs+01,fnswbpr02,fss++03,wbfs03}. 
This implies that the GP optical depth
at $z \gtrsim 6$ is larger than a few. Unfortunately, such a constraint
does not necessarily imply that the universe is neutral at such redshifts.
For example, a neutral hydrogen fraction $\bar{x}_{\rm HI} \sim 10^{-3}$
would produce an optical depth $\tau_{\rm GP} \sim 100$, more than 
what is required
to produce the absorption troughs. 
This is the typical level of constraint one can obtain through
such model-independent simplistic arguments based on an uniform medium.
Such arguments, though quite effective in giving robust conclusions
at low redshifts, do not yield   any strong constraint on the 
neutral hydrogen fraction at $z \gtrsim 6$.

The next line of argument for the approach to the final stages
of reionization at $z \gtrsim 6$
is based on the change in the slope of the optical
depth \cite{fnswbpr02,cm02,wbfs03}
around $z \sim 5.5 - 6$, which indicates that some qualitative change in the
physics of IGM occurs at these redshifts.
To understand this in simple terms, let us write the neutral hydrogen
fraction $\bar{x}_{\rm HI}$ in terms of more physically 
meaningful quantities:
\begin{equation}
\bar{x}_{\rm HI} = 2.7 \times 10^{-5} 
\left(\frac{T_0(z)}{10^4 {\rm K}}\right)^{-0.7}
\left(\frac{\Gamma_{\rm PI}(z)}{10^{-12} {\rm s}^{-1}}\right)^{-1}
\left(\frac{\Omega_b h^2}{0.022}\right)
\left(\frac{1-Y}{0.76}\right)
\left(\frac{1+z}{7}\right)^3
\label{eq:x_hi}
\end{equation}
where $T_0$ is the temperature of the medium at the mean density 
($\Delta =1$) and $\Gamma_{\rm PI}$ is the photoionization rate of neutral
hydrogen (assumed to be homogeneous).
Combining the above equation with (\ref{eq:tau_gp}), one can see that
$\tau_{\rm GP}(z) \propto (1+z)^{4.5} T_0^{-0.7}(z)/\Gamma_{\rm PI}(z)$.
Thus when $T_0$ and $\Gamma_{\rm PI}$ are not changing 
substantially with redshift, we expect 
$\tau_{\rm GP}(z) \propto (1+z)^{\alpha}$ with $\alpha \approx 4.5$.
This is indeed seen in the observations at $z \lesssim 5.5$ \cite{fsb++05}.
However, at higher redshifts, the observations show that 
$\tau_{\rm GP}$ evolves much faster combined with a rapid deviation
from a power-law evolution, thus implying that the properties
of IGM (like $T_0$ and $\Gamma_{\rm PI}$) are evolving considerably.
This argument points towards a possible phase change in the IGM and 
thus suggesting that we are approaching the final stages 
of reionization at $z \approx 6$.
However, one should keep in mind that this argument does {\it not}
conclusively prove that the IGM is neutral at $z \gtrsim 6$ -- it
simply indicates for a rapid change in properties. Furthermore, the 
above defined $\tau_{\rm GP}$ is {\it not} a directly measured
quantity; one instead measures the mean transmitted flux $\bar{F}$ which
is computed by integrating the optical depth over all possible 
overdensities:
\begin{equation}
\bar{F} = \int_0^{\infty} {\rm d} \Delta ~ P(\Delta) ~ 
\exp[-\tau_{\rm GP}(\Delta)].
\end{equation}
The quantity $P(\Delta)$ denotes the density distribution of the IGM.
It is thus clear from the above expression that any robust conclusion
based on the observed evolution of 
$\bar{F}$ would require a good knowledge of $P(\Delta)$.

Hence, the next step for calculating the ionization properties
of the IGM from QSO spectra is to include the density inhomogeneities
in the analysis. From this point on, the conclusions become extremely model-dependent as
we do not have a clear understanding of the density distribution
of the IGM.  
One approach would be to use numerical simulations 
for obtaining the IGM density distribution and
then compute the absorption spectra of high-redshift quasars in the 
Ly$\alpha$ region \cite{fnswbpr02}. Using this approach, 
a rapid evolution of the 
volume-averaged neutral fraction of hydrogen has been found at $z \lesssim 6$
($\bar{x}_{\rm HI} \sim 10^{-5}$ at $z = 3$ to 
$\bar{x}_{\rm HI} \sim 10^{-3}$ at $z = 6$). On the other hand, 
a different set of analyses \cite{sc02,songaila04}
from nearly similar data set conclude that
the transmitted fractions have a relatively smooth evolution 
over the entire range of redshifts, which can be modeled with a 
smoothly decreasing ionization rate; hence no evidence of a 
rapid transition could be established.

In addition to the global statistics discussed above, 
there are some results based on the transmission
observed in the spectra of individual sources. For example, the
analyses of the spectrum of the most distant known quasar (SDSS
J1148+5251) at an emission redshift of 6.37 
show some residual flux both in the Ly$\alpha$ and
Ly$\beta$ troughs, which when combined with Ly$\gamma$ region
\cite{fo05}, imply that this flux is consistent with pure
transmission.  The presence of unabsorbed regions in the spectrum
corresponds to a highly ionized IGM along that particular line of
sight.  However, a complete GP
trough was detected in the spectrum of SDSS J1030+0524 ($z=6.28$) 
\cite{bfw++01}, where no
transmitted flux is detected over a large region (300 \AA) immediately
blueward of the Ly$\alpha$ emission line.  Such differences in the 
ionization state of the IGM along different lines of sight have been 
interpreted as a possible signature of the pre-overlap phase of reionization.

There have been other different approaches to investigate the neutral
hydrogen fraction. For example, one can estimate the
sizes of the ionized regions around the QSOs from the spectra 
\cite{wl04,wlc05}. 
Then the neutral gas
surrounding the QSO can be modelled as a function of different parameters: the
Str\"omgren sphere size $R_S$, the production rate of
ionizing photons $\dot{N}_{\rm ph}$ from the QSO, 
the clumping factor of the gas
$C$ and the age of the QSO $t_{\rm age}$. 
Considering 7 QSOs at $z > 6$ (which included
the above cited QSOs), it has been argued that 
the small sizes of the ionized regions ($\sim 10$ physical Mpc)
imply that the typical neutral hydrogen fraction of the IGM beyond
$z\sim 6$ is in the range 0.1 - 1.  However, this approach is weighted
down by several uncertainties.  For example, one of the uncertainties
is the quasar's production rate of ionizing photons $\dot{N}_{\rm ph}$ 
as it depends on the shape of the spectral template used.
Moreover it is implicitly assumed in the modelling of clumping factor
that the formation of quasars and galaxies were simultaneous. This in
turn implies that quasars ionize only low density regions and hence
the clumping factor, which regulates the evolution of the ionized regions,
is low. If, instead, stars appears much earlier than QSOs, the quasars
have to ionize high density regions, which means that one should use a
higher value of clumping factor in the calculations \cite{yl05}.

There has been a different approach based on the damping wings
of the neutral hydrogen \cite{mh04}. 
Using density and velocity fields obtained by
hydrodynamical simulation, the Ly$\alpha$ absorption spectrum 
was computed. In this case the neutral hydrogen fraction,
$\dot{N}_{\rm ph}$ and $R_S$ are treated as free parameters,
constrained by matching the optical depth observed in the QSO SDSS
J1030+0524. Also in this case, the conclusion is that 
the neutral hydrogen fraction is
larger than 10 per cent, i.e., the IGM is significantly more neutral at
$z\sim 6$ than the lower limit directly obtainable from the GP trough
of the QSO spectrum ($\bar{x}_{\rm HI} \approx 10^{-3}$).  However this result is based only on
one quasar. Moreover, the observational constraints on the optical
depth are very uncertain and can introduce errors in the estimates of
$\bar{x}_{\rm HI}$.

To summarise the QSO absorption line observations -- 
there is still {\it no} robust and model-independent 
constraint on the neutral hydrogen fraction from the data.
The spectroscopy
of the Ly$\alpha$ forest for QSOs at $z > 6$ discovered by the 
SDSS \cite{fnl+01,fss++03} 
strongly suggest that the IGM is highly ionized along some lines of sight.
On the other hand, there are a few (maybe a couple) lines of sight
which seems to indicate that the IGM is neutral, though the
conclusion is still not robust.
In case we find transmission along some lines of sight while the
medium seems quite neutral along others could possibly be interpreted
that the IGM ionization properties are different along different lines
of sight at $z \gtrsim 6$, thus suggesting that we might be observing
the end of the reionization process. However, it is also possible that
such dispersion in the IGM properties along different lines of sight
can be accommodated by simply the dispersion in the density inhomogeneities.
As discoveries of more such objects are expected in future,
spectroscopy of high-redshift QSOs remains one of the principal empirical 
approaches to understand the final stages of reionization.

Before completing our discussion on the QSO absorption lines, it is
worth mentioning a set of indirect constraints on reionization based on the
temperature of the IGM at $z \approx 2-4$.\footnote{This
determination of temperature puts constraints on the 
reionization of helium too; however, the helium reionization 
is beyond the scope of
this review.} Using various techniques
like, the lower envelope of the neutral hydrogen column density and 
velocity width scatter plot \cite{stres00,rgs00} or wavelet transforms
\cite{tszktc02}, one can infer the 
temperature of the IGM from absorption lines. These analyses
suggest that $T_0 \sim 1-2 \times 10^4$ K at $z \approx 3$, which 
in turn imply that hydrogen reionization must occur at $z < 9$ or else
the temperature would be too low to match the observations. However, 
one should keep in mind that the analyses has large uncertainties, 
like, for example, the dust photoheating of the IGM
could give rise to high temperatures at $z \approx 3$ \cite{fnss99,nss99,ik03,ik04}. 
Furthermore, 
a complex ionization history of helium could relax considerably
the constraints obtained from $T_0$
on the reionization epoch.

\subsubsection{Cosmic microwave background radiation}

The second most important analysis regarding the reionization
history comes from the observations of temperature and polarization
anisotropies in the cosmic microwave background (CMB) radiation.
As far as the temperature anisotropies are concerned, reionization 
can damp the fluctuations on small scales due to photon diffusion 
in the ionized plasma. The scattering of photons suppresses the 
anisotropies on angular scales below the horizon at the rescattering epoch 
by a damping factor ${\rm e}^{-\tau_{\rm el}}$, where 
\begin{equation}
\tau_{\rm el} = \sigma_T c \int {\rm d} t ~ n_e ~ 
(1+z)^3 
\end{equation}
is the optical depth (measured at the present epoch) 
of CMB photons due to Thomson scattering 
with free electrons. In the above expression, $n_e$ is the 
average value of the comoving electron density and $\sigma_T$ is 
the Thomson scattering cross section. 
However, measuring this damping is not easy as it 
can be compensated by a larger strength of dark matter density
fluctuations which are measured by the 
corresponding power spectrum, usually parametrised
by the two quantities: the primordial spectral index $n_s$ and 
the fluctuation amplitude at cluster scales $\sigma_8$. Hence, it is found that
$\tau_{\rm el}$ is only mildly constrained by the temperature fluctuations
because of strong degeneracies with $n_s$ and 
$\sigma_8$ \cite{eht98}. For this reason, 
temperature anisotropy data prior to Wilkinson Microwave Anisotropy Probe 
(WMAP)\footnote{http://map.gsfc.nasa.gov/} could only constrain 
$\tau_{\rm el} \lesssim 0.5$ \cite{saa++01}.
For sudden reionization models, this only implies that
the redshift of reionization 
$z_{\rm re} \lesssim 40$. To put in perspective with the discussion
of QSO absorption line observations, reionization at $z \sim 6$ would
imply $\tau_{\rm el} \sim 0.05$.

A major breakthrough in our understanding of reionization came after
the release of first year WMAP results of polarization measurements.
A fundamental prediction of the gravitational instability paradigm is
that CMB anisotropies are polarized, i.e., if the temperature
anisotropies are produced by primordial fluctuations, their presence
at the last scattering surface would polarize the CMB. The generation
of polarization requires two conditions to be satisfied: (i) photons
need to undergo Thomson scattering off free electrons (the
corresponding cross section is polarization-dependent) and (ii) the
angular distribution of the photon temperature must have a non zero
quadrupole moment.  Tight coupling between photons and electrons prior
to recombination makes the photon temperature almost isotropic and the
generated quadrupole anisotropy, and hence the 
polarization, is very small. Because the
temperature anisotropies are of the order $10^{-5}$, the polarization
is about $10^{-6}$ or less. 

To generate a quadrupole, it is
necessary to produce velocity gradients in the 
photon-baryon fluid across
the photon mean free path; hence only those
perturbations which have length scales smaller than
the mean free path can produce polarization. At larger scales, 
multiple scattering will make the plasma quite homogeneous and thus
no significant quadrupole can be generated, while at much 
lower scales polarization is suppressed
due to ``Silk damping''.
In fact, the polarization generated at the last scattering surface
would be significant at scales comparable to
the horizon size at that epoch (which corresponds
to a multipole number $\ell \sim 100$), and 
{\it no} polarization signal is expected at larger scales $\ell < 100$.
Detection of polarization signal at $\ell < 100$ is a clear signature
of secondary processes such as reionization.

Following the completion of 
recombination the quadrupole moment of temperature anisotropies 
grows due to the
photons free streaming. In case these photons are able
to scatter off free electrons at a later stage,
the anisotropy can be transformed into substantial
polarization. This is an ideal effect to probe reionization as it is
the only process which can provide considerable number of free electrons
at post-recombination epochs.
For models with sudden reionization, it can be shown  
that the effect dominates on the angular scale of the
horizon at the epoch of reionization. The polarization signal 
will peak at a position $\ell \propto
z_{\rm re}^{1/2}$ 
 with an amplitude proportional to
the total optical depth $\tau_{\rm el}$. 
Thus the polarization spectrum at low $\ell$ is
a sensitive probe of the reionization process.

The polarization measurements by
the WMAP satellite \cite{ksb++03} found a significant signal in
the temperature-polarization cross correlation spectrum at $\ell <
10$. The position and the amplitude of this excess 
is consistent with an optical depth $\tau_{\rm el} = 0.17
\pm 0.04$, implying a (sudden) reionization redshift $11 < z_{\rm re}
< 30$. While this result has 
possibly complicated the picture of reionization and thus 
generated tremendous amount of activity
within the community, a few subtleties should be kept in mind while
using the reionization constraints: (i) The result is based 
on a few points at low $\ell$ 
and it is necessary that such an important result  is confirmed by future data.
One should also note that the likelihood function for $\tau_{\rm el}$ 
obtained from WMAP data is heavily skewed, probably indicating some sort
of a ``tension'' within the data.
(ii) The constraints on $\tau_{\rm el}$ depend on the priors and
analysis technique used. For example, fitting the temperature -- E-mode
polarization cross power
spectrum (TE) to $\Lambda$CDM models in which all parameters except
$\tau_{\rm el}$ assume their best fit values based on the temperature power
spectrum (TT), the 68\% confidence range obtained is
$0.13<\tau_{\rm el}<0.21$ \cite{ksb++03}. Fitting all
parameters simultaneously to the TT and the TE data, 
the corresponding range changes to 
obtain $0.095<\tau_{\rm el}< 0.24$ \cite{svp++03}.  
Including additional data
external to WMAP, these authors were able to shrink their confidence
interval to $0.11<\tau_{\rm el}<0.23$.  Finally, by assuming that the
{\it observed} TT power spectrum is scattered to produce the observed
TE cross-power spectrum, the inferred range is $0.12<\tau_{\rm
  el}<0.20$ \cite{ksb++03}.
(iii) The constraints of $\tau \approx 0.17$ and $z_{\rm re} 
\approx 15$ usually quoted in the literature assume a sudden reionization.
The constraints can change drastically when this assumption is relaxed.

In case the result is confirmed by future data sets,
we note
that it is not necessarily contradictory to the QSO results; the history of
the luminous sources and their effect on the IGM was probably highly complex,
and there was a finite time interval 
(maybe somewhere around a few hundred million years) 
from the appearance of the first sources
of UV photons and the completion of the
reionization.

\subsubsection{Ly$\alpha$ emitters at high redshifts}

In parallel, a number of groups have studied star-forming galaxies at
$z \sim 6 - 7$, and measurements of the Ly$\alpha$ emission line luminosity
function evolution provide another useful observational constraint
\cite{mr04,sye++05}.
While the QSO absorption
spectra probe the neutral hydrogen fraction regime $x_{\rm HI} \leq 0.01$,
this method is sensitive to the range $x_{\rm HI} \sim 0.1 - 1.0$.
Ly$\alpha$ emission from galaxies is expected to be 
suppressed at redshifts beyond reionization because of the 
absorption due to neutral hydrogen, which clearly 
affects the evolution of the luminosity function of such 
Ly$\alpha$ emitters
at high redshifts \cite{hc05,mr04,fhz04}.
Thus a comparison of the luminosity functions
at different redshifts could be used for constraining the reionization.
Through a simple analysis, it was found that the luminosity functions 
at $z = 5.7$  and $z = 6.5$ are statistically consistent with one another,
thus implying that reionization was largely complete at $z \approx 6.5$.
More sophisticated calculations on the evolution of the luminosity function of Ly$\alpha$
emitters \cite{mr04,fzh06,hc05} suggest that the neutral fraction
of hydrogen at $z=6.5$ should be less than 50 per cent
\cite{mr05}.

The analysis of the  Ly$\alpha$ emitters at high redshifts is complicated
by various factors.
(i) Firstly, this suppression of the Ly$\alpha$ emission line 
depends on the size of the ionized 
region surrounding the source as larger ionized volumes allow
more photons to escape.
On the other hand, 
the sizes of the ionized regions themselves depend on the clustering 
properties of the sources. 
There is thus a strong coupling between
the clustering of the sources,  
sizes of the ionized regions and the luminosity function of the 
Ly$\alpha$ emitters at high redshifts. 
(ii) The ionized hydrogen regions are typically highly asymmetric because
the the ionization-fronts propagate much faster across underdense voids then
across dense filaments. Thus one needs to know the details of the 
density distribution around the sources to model the ionized regions.
(iii) It is well known that bright galaxies are biased, so it is likely
that more than one galaxy is located inside a single ionized region;
Ly$\alpha$-emitters can also be located inside ionized regions of luminous
quasars, which are often many times larger than the ionized regions of
galaxies. It is thus clear that the modelling of the ionized regions 
of Ly$\alpha$-emitters is not straightforward, and hence the reionization
constraints could be severely model-dependent.

\subsection{Observations related to the sources of reionization}

As we discussed in the Introduction, a major challenge 
in our understanding of reionization depends on our knowledge
of the sources, particularly at high redshifts. In this sense, 
reionization is closely related to formation of early baryonic structures
and thus any observation related to the detection of very distant sources
can be important for constraining reionization. In the following, we 
shall discuss a few most important of such observational probes.

\subsubsection{Direct observations of sources at high redshifts}

As we understand at present, neither the bright
$z> 6$ QSOs discovered by the SDSS group \cite{fsr++05} nor the
faint AGN detected in X-ray observations \cite{bcc++03} 
produce enough photons to reionize the IGM. 
The discovery of star-forming galaxies
at $z>6.5$ \cite{hcm++02,ktk++03,kesr04}
has resulted in speculation that
early galaxies produce bulk of the ionizing photons for reionization. 
However, the 
spectroscopic studies of I-band dropouts in the Hubble
ultra-deep Field 
with confirmed redshifts at $z \approx 6$, indicate
that the measured star formation rate at $z = 6$ is lower by factor of 6 from
the $z = 3$ star formation rate. 
If the estimate is correct, the I-dropouts do not
emit enough ionizing photons to reionize the universe at $z \approx 6$
\cite{bsem04}. The
short-fall in ionizing photons might be alleviated 
by a steep faint-end slope of the luminosity
function of galaxies or a different stellar initial
mass function (IMF); alternatively, the bulk of reionization might 
have occurred at $z > 6$ through rapid star formation in galaxies at much
higher redshifts.

There are estimates of a somewhat higher UV luminosity at 
$z = 6-10$. This is obtained by 
constructing a luminosity function from $\sim 500$ galaxies collected from
all the deepest wide-area HST data \cite{bi05}. 
The luminosity function thus obtained
extends 3 magnitudes fainter than the characteristic luminosity $L^*$.
This analysis predicts a significant evolution in $L^*$ --­ a doubling from 
$z = 3$ to
$z = 6$, thus  implying
a luminosity density that is only a factor of 1.5 less than
the luminosity density at $z = 3$. 
The observed evolution is suggestive of that expected 
from popular hierarchical models, and
would seem to indicate that we are literally witnessing the 
buildup of galaxies in the reionization era.

To summarise, there are somewhat conflicting reports regarding the star 
formation rate at 
$z \gtrsim 6$ -- however, it is safe to conclude that we have not yet
observed enough number of sources which could ionize the bulk
of the IGM at $z \gtrsim 6$.
Whether the reionization was actually completed by 
galaxies at a much higher redshifts is still an open
issue.

\subsubsection{Cosmic infrared background radiation}

Numerous arguments favour an excess contribution to the extragalactic 
background light between 1 $\mu$m and a few $\mu$m 
\cite{mcf++00,gwc00,crbj01,wright01}
when compared to the expectation based on galaxy counts and 
Milky Way faint star counts (for a review see \cite{hd01}). 
While these measurements are likely to be affected by 
certain systematics and issues related to the exact contribution 
from zodiacal light within
the Solar System, one explanation is that a contribution to the 
cosmic infrared background (CIRB) radiation originates
from high redshift sources. The redshifted line emission from 
Ly$\alpha$ emitting galaxies at $z > 9$ would produce an
integrated background in the near-infrared
wavelengths observed today. In case this interpretation of the CIRB is
correct, it would directly constrain the number of ionizing sources
at high redshifts and thus would have direct implications on reionization.

However, if the entire CIRB is due to the high redshift
galaxies, the explanation requires the presence of metal-free
PopIII stars with a top-heavy IMF and possibly a high star-forming 
efficiency \cite{sf03,msf03,kagmm04,cbklm04,ms05}. 
In fact, the number of sources required to explain the 
CIRB is much higher than that needed to explain the early reionization
constraints. The most serious difficulty in explaining the 
CIRB through PopIII stars comes from the observations 
of the number of J-dropouts and Ly$\alpha$ 
emitters in ultra deep field searches as the models severely 
overpredict the number of sources \cite{sf05}. At present, the origin 
of CIRB remains to be puzzling (as one can discard
other possible sources like miniquasars and decaying neutrinos, see the
next subsection), 
and it is not clear whether it could have any significant implications on reionization.


\subsubsection{Constraints on other sources}

We end this section by briefly reviewing the constraints we have 
on other kind of sources, namely the Intermediate Mass Black Holes and
decaying (exotic) particles.

A large population of intermediate mass black holes (IMBHs) might be
produced at early cosmic times as a left over of the evolution of 
very massive first stars. These  black holes at high redshifts ($z>6$) 
can, in principle, contribute to the ionization of the 
IGM; however they would be accompanied by the 
copious production of hard X-ray photons (with energies above
10 keV). The resulting hard X-ray background would redshift and be 
observed as a present-day soft X-ray background. 
One can show that the observed
residual soft X-ray background intensity can put stringent
constraints on the the baryon mass fraction locked
into IMBHs and their growth \cite{dhl04,shf05}. Thus, unless they are
extremely X-ray quiet, these black holes, or miniquasars, 
must be quite rare and/or have a
short shining phase. As a byproduct, it implies that miniquasars cannot
be the only source of reionization.

The other sources which are popularly invoked to explain reionization
are the exotic particles like decaying neutrinos \cite{sciama88,dj92,dj94,sciama94,sciama95,sn97}. However, in most
cases these particles 
decay radiatively (producing photons) and hence are severely 
constrained by Big Bang Nucleosynthesis, diffuse soft X-ray and gamma-ray backgrounds 
and the deviation 
of the CMB spectrum from Planckian shape.
For example, the constraints from soft X-ray background limits the 
radiatively decaying sterile neutrino mass to $m_{\nu} < 14$ keV and
hence the optical depth to Thomson scattering is $\tau_{\rm el} \sim 10^{-2}$,
negligible compared to what is required for explaining observations
\cite{mf05}. 
Similar constraints exist for other particles, including those 
which have decay channels into electrons instead of photons \cite{hh04}.
The point what comes out from most these analyses is that  
different observational constraints leave out a very
small parameter space accessible to the decaying particles and hence their
contribution to reionization may not be that significant.

\section{Physics of reionization}

Given the observational constraints discussed in the previous Section, it is important to develop models which can be reconciled with every data set. However, such a task is not straightforward simply because of the complexities in the physical processes involved. It is {\it not} that there is some unknown physics involved -- we believe that we can write down every relevant equation -- the difficulty lies in solving them in full generality. This is quite a common obstacle in most aspects of large scale structure formation, but let us concentrate on reionization for the moment.

A crucial issue about reionization is that this process is tightly
coupled to the properties and evolution of star-forming galaxies and
QSOs. Hence, the first requirement that any reionization model
should fulfill is that it should be able to reproduce the available
constraints concerning the luminous sources. Though there has been a tremendous
progress in our understanding of formation of galaxies and stars at low redshifts, very little is known about the high redshift galaxies, particularly those belonging to the first generation. 
There are strong indications, both from numerical simulations and analytical
arguments, that the first generation stars were metal-free, and hence
massive, with a very different kind of spectrum than the stars 
we observe today \cite{schaerer02}; they are known as the PopIII stars.
Along similar lines, 
whereas a relatively
solid consensus has been reached on the luminosity function, spectra
and evolutionary properties of intermediate redshift QSOs, some
debate remains on the presence of yet undetected low-luminosity
QSOs powered by intermediate mass black holes at high redshift
\cite{ro04b}.  Even
after modelling the ionizing sources to a reasonable degree of accuracy, 
predicting the joint reionization and star formation
histories self-consistently is not an easy task as 
(i) we still do not have a clear idea on how the photons ``escape'' from 
the host galaxy to the surrounding IGM, and 
(ii) mechanical and
radiative feedback processes can alter the hierarchical structure
formation sequence of the underlying dark matter distribution as far
as baryonic matter is concerned.

A different approach to studying reionization would be to parametrize the 
sources by various adjustable parameters (like efficiency of star formation, 
fraction of photons which can escape to the IGM, etc) and then
calculate the evolution of global 
ionization and thermal properties of the IGM. Though it is possible
to obtain a good qualitative picture of reionization through this approach, it is difficult to obtain the details, particularly those of the 
pre-overlap phase, through such simple
analyses. For example, many 
details of the reionization process can be dealt 
only in an approximate manner (the
shape of ionized region around sources and their overlapping, just to
mention a few) and in terms of global averages (such as the filling
factor and the clumping factor of ionized regions).
We shall discuss this in more detail later in this Section.

It is thus clear that the complexities within the physics of reionization
prohibit us from constructing detailed analytical models. The other option
is to solve the relevant equations numerically and follow the ionization
history.  However, one should realize
that in order to exploit the full power of the observational
data available to constrain models, one must be able to connect widely
differing spatial and temporal scales. In fact, it is necessary, at
the same time, to resolve the IGM inhomogeneities (sub-kpc physical
scales), follow the formation of the very first objects in which
stars form (kpc), the radiative transfer of their ionizing
radiation and the transport of their metals (tens of kpc), the build
up of various radiation backgrounds (Mpc), and the effects of QSOs and 
sizes of the ionized regions (tens of
Mpc). Thus, a proper modelling of the relevant physics on these
scales, which would enable a direct and {\it simultaneous} comparison
with all the available data set mentioned above, would require
numerical simulations with a dynamical range of more than five orders of
magnitude, which is far cry from the reach of our current
computational technology. To overcome the problem, simulations have
typically concentrated on trying to explain one (or few, in the best
cases) of the observational constraints. It is therefore difficult
from these studies to understand the extent to which their conclusions
do not conflict with a different set of experiments other than the one
they are considering.

However, one must realise that in spite of these 
difficulties in modelling reionization
there have been great progresses in recent years, 
both analytically and 
through numerical simulations, in different aspects of the process. 
This Section will be devoted to the successes we have achieved towards
understanding reionization.

\subsection{Analytical approaches}

As far as the analytical studies are concerned, there are two broad 
approaches in modelling, namely, (i) the evolution of ionized 
regions of individual ionizing sources and (ii) statistical approaches 
in computing the globally averaged properties and fluctuations. 

\subsubsection{Evolution of ionized 
regions of individual ionizing sources}

The standard picture of reionization by discrete sources of radiation
is characterized by the expansion and overlap of  the individual ionized 
regions. In this paradigm, it is important to understand how the 
ionized regions evolve for different types of ionizing sources.

The most common sources studied are the ones with UV photons,
i.e., photons with energies larger than 13.6 eV but within few tens of eV.
These photons ionize and heat up the IGM through photoionizing neutral 
hydrogen (and possibly helium). 
Because of (i) a large value of the photoionization cross section around
13.6 eV, (ii) rapid increase of the number of absorbers 
(Ly$\alpha$ ``clouds'') with lookback time 
and (iii) severe attenuation of sources at higher redshifts, 
the mean free path of photons at 13.6 eV becomes so small beyond a redshift
of 2 that the radiation is largely ``local''. For example, the 
mean free path at 13.6 eV is typically $\approx 30$ proper Mpc at 
$z \approx 3$, which is much smaller than the horizon size.
In this approximation, the background intensity 
depends only on the instantaneous value of the 
emissivity (and not its history) because all the photons are
absorbed shortly after being emitted (unless the 
sources evolve synchronously over a timescale much
shorter than the Hubble time).

  When an isolated source of ionizing radiation, say a star or a QSO, turns
on, the ionized volume initially grows in size at a rate determined 
by the emission of UV photons. The boundary of this volume
is characterised by an ionization front
which separates the ionized and neutral regions and propagates into the
neutral gas. Most photons travel freely in the ionized
bubble and are absorbed in a transition layer. 
Across the front the
ionization fraction changes sharply on a distance of the
order of the mean free path of an ionizing photon (which is much 
smaller than the horizon scale). 
The evolution of an expanding ionized region is governed by the equation 
\cite{sg87,mhr99}
\begin{equation}
\frac{{\rm d} V_I}{{\rm d} t}
= \frac{\dot{N}_{\rm ph}}{n_{\rm HI}} 
- \frac{V_I}{t_{\rm rec}} 
\label{eq:stromgren}
\end{equation}
where $V_I$ is the comoving volume of the ionized region, $\dot{N}_{\rm ph}$
is the number of ionizing photons
emitted by the central source per unit time, $n_{\rm HI}$ is the mean 
comoving density of neutral hydrogen and
$t_{\rm rec}$ is the recombination timescale of neutral hydrogen given by
\begin{equation}
t_{\rm rec}^{-1} = {\cal C} \alpha_R n_{\rm HI} (1 + z)^3 
\end{equation}
In the above relation $\alpha_R$ denotes the recombination rate 
ionized hydrogen and free electrons and
${\cal C}$ is the clumping factor which takes into account the 
enhancement in the number of recombinations due to density inhomogeneities.
It is clear from equation (\ref{eq:stromgren}) that 
the growth of the ionized bubble is slowed down by
recombinations in the highly inhomogeneous IGM. 
In the (over)simplified case of both $\dot{N}_{\rm ph}$ and
$t_{\rm rec}$ not evolving with time, 
the evolution equation can be solved exactly to obtain
\begin{equation}
V_I = \frac{\dot{N}_{\rm ph}~t_{\rm rec}}{n_{\rm HI}} 
\left(1 - {\rm e}^{-t/t_{\rm rec}}\right) 
\end{equation}
While the volume of the ionized region depends on the
luminosity of the central source, the time it takes to produce
an ionization-bounded region is only a function of $t_{\rm rec}$. 
It is clear that the ionized volume will approach its Str\"omgren
radius after a few recombination timescales.
Thus, when
the recombination timescale is similar to or larger than the Hubble time 
(which is usually the case at lower redshifts, but depends 
crucially on the value of the clumping factor),
the ionized region will never reach the Str\"omgren radius.
The solution of the equation is more complicated in reality where
the sources evolve, but the qualitative feature remains the same. 
After the bubbles have grown sufficiently, these individual 
bubbles start overlapping with each other eventually complete the 
reionization process.

As far as the hydrogen reionization is concerned, the evolution of the 
ionization front for stars and QSOs are qualitatively similar. 
The only difference is that the QSOs are much more luminous than
stellar sources and hence the ionization front propagates much 
faster for QSOs. 
In the case of doubly-ionized helium, however,
the propagation of ionization fronts for the two types 
of sources differ drastically.\footnote{The propagation of
singly-ionized helium front coincides with the 
hydrogen ionization front for almost all forms of 
the ionizing spectrum. Hence, as far as helium is concerned, most
studies are concerned with the propagation of the doubly-ionized helium
fronts.} Since normal stars
hardly produce any photons above 54.4 eV, the propagation of the 
doubly-ionized helium ionization front is negligible; 
in contrast the QSOs have
a stiff spectrum and one can show that the doubly-ionized helium front 
not only propagates quite fast into the medium, but closely
follows the ionized hydrogen one. The analyses actually predict that 
the ionization of hydrogen
and double-ionization of helium
would be nearly simultaneous if QSOs were
the dominant source of reionization, which cannot be the 
case for normal stellar population. Also note that the early metal-free 
massive PopIII stars too have a hard spectrum, and in their case the
reionization of 
hydrogen and singly-ionized helium is simultaneous, similar to QSOs.
This could provide an indirect way of identifying the sources of reionization
from future observations.

If the sources of radiation produce photons of much higher energies, say
in the X-ray band, the nature of the ionization front changes considerably.
The most common example of such sources is the early population of 
accreting black holes, or miniquasars \cite{vgs01,oh01}. 
These sources are expected
to produce photons in the X-ray bands. The photons which are most
relevant for this discussion would be those with energies below 2 keV, 
as photons with higher energies have mean free path similar to
the horizon scale and
thus would rarely be absorbed. 

Since the absorption cross section of neutral hydrogen varies 
with frequency approximately
as $\nu^{-3}$, the mean free path for photons with high energies 
would be very large. A simple calculation will show that for photons
with energies above 100--200 eV, the mean free path would be larger
than the typical separation between collapsed structures \cite{mrvho04}
(the details
would depend upon the redshift and exact description of collapsed 
haloes). These photons would not be associated with any particular
source at the moment when they are absorbed, and thus would
ionize the IGM in a more homogeneous manner (as opposed to the 
overlapping bubble picture for UV sources). 
However, one should also
note that the number of photons produced in the X-ray bands
is usually not adequate for fully ionizing the IGM. 
Hence these hard photons can only partially
ionize the IGM through repeated secondary ionizations. Basically, the
photoelectron produced by the primary ionization would be very energetic
($\sim 1$ keV) and could thus produce quite a few ($\sim 10$)
secondary electrons via collisional ionization. This may not be a very
effective way of producing free electrons (when compared
to the UV background), but is an efficient way of heating up the medium.
This very population of hard photons deposits a fraction of 
its energy and can heat up the IGM
to $\sim 100 - 1000$ K \cite{cm04} before reionization -- a process sometimes known
as preheating.

In case of reionization through more exotic sources like decaying particles,
the nature of ionized volumes would be completely different. The production
of ionizing photons (or electrons) in this case is not related to any 
collapsed structure -- rather it occurs throughout the space in an
homogeneous manner. In such case, one expects only limited  
patchiness in the distribution of the background radiation or ionized fraction.

We have discussed various different ways in which the ionized 
regions could evolve depending on the nature of the ionizing sources.
The next step would be to take into account the global distribution 
of the sources and ionized volumes and thus construct the 
global  picture of reionization, which we shall study in the 
next subsection.

\subsubsection{Statistical approaches}

The most straightforward statistical quantity which is studied in 
reionization is the volume filling factor of ionized regions $Q_I$. 
For the most common UV sources, this is obtained by averaging
equation (\ref{eq:stromgren}) over a large volume:
\begin{equation}
\frac{{\rm d} Q_I}{{\rm d} t}
= \frac{\dot{n}_{\rm ph}}{n_{\rm HI}} 
- \frac{Q_I}{t_{\rm rec}} 
\end{equation}
where $\dot{n}_{\rm ph}$ is average number of ionizing photons
produced per unit volume per unit time. Note that the above
equation implicitly assumes that the sources of radiation are
distributed uniformly over the volume we are considering. The equation
can be solved once we know the value and evolution of the photon
production rate $\dot{n}_{\rm ph}$ and also the clumping factor ${\cal C}$. 
One can get an estimate of $\dot{n}_{\rm ph}$ 
by assuming that it is proportional to 
the fraction of gas within collapsed haloes, while the value of 
${\cal C}$ is relatively difficult to calculate and is usually 
assumed to be somewhere between a few and 100. In this simple
picture, reionization is said to be complete when $Q_I$ reaches unity.
 Assuming 
a normal stellar population (i.e., a Salpeter-like IMF and standard
stellar spectra obtained from population synthesis codes) and a
value of ${\cal C} \approx 10-30$, one obtains a reionization
in the redshift range between 6 and 10 \cite{lb01}.

This approach can be improved if the density inhomogeneities of the IGM are taken into account.
In the above picture, the inhomogeneities in the IGM are considered
simply in terms of the clumping factor in the effective recombination
timescale without taking into account the density distribution of the IGM.
The importance of using a density distribution of the IGM lies in the
fact that regions of lower densities will be ionized first, and
high-density regions will remain neutral for a longer time. The main
reason for this is that the recombination rate is higher in
high-density regions where the gas becomes neutral very 
quickly\footnote{Of
course, there will be a dependence on how far the high density region
is from an ionizing source; however such complexities can only be dealt
in a full numerical simulation.}. 
Thus, in the situation where all the individual ionized
regions have overlapped (the so-called post-overlap stage), 
all the low-density regions (with overdensities, say, $\Delta < \Delta_i$)
will be highly ionized, while there will be some high
density peaks (like the collapsed systems) which will still remain
neutral. The situation is slightly more complicated when the ionized regions
are in the pre-overlap stage. At this stage, it is assumed that a
volume fraction $1 - Q_I$ of the universe is completely neutral
(irrespective of the density), while the remaining $Q_I$ fraction of
the volume is occupied by ionized regions. However, within this
ionized volume, the high density regions (with $\Delta > \Delta_I$)
will still be neutral. Once $Q_I$ becomes unity, all regions with
$\Delta < \Delta_I$ are ionized and the rest are neutral. The
high-density neutral regions manifest themselves as the Lyman-limit
systems (i.e., systems with neutral hydrogen column densities 
$N_{\rm HI} > 10^{17}$ cm$^{-2}$)
in the QSO absorption spectra. The reionization process after
this stage is characterized by increase in $\Delta_I$ -- implying that
higher density regions are getting ionized gradually \cite{mhr00}.

To develop the equations embedding the above physical picture, we need
to know the probability distribution function $P(\Delta) {\rm d} \Delta$
for the overdensities. Given a $P(\Delta) {\rm d} \Delta$, it is clear that
only a mass fraction
\begin{equation}
F_M(\Delta_I) = \int_0^{\Delta_I} {\rm d} \Delta ~ \Delta ~ P(\Delta)
\end{equation}
needs be ionized, while the remaining high density regions will be
completely neutral because of high recombination rates. The
generalization of equation (\ref{eq:stromgren}), appropriate for this
description is given by \cite{mhr00,wl03}
\begin{equation}
\frac{{\rm d} [Q_I F_M(\Delta_I)]}{{\rm d} t} = 
\frac{\dot{n}_{\rm ph}(z)}{n_{\rm HI}} 
- Q_I \alpha_R n_{\rm HI} R(\Delta_I) (1 + z)^3
\label{eq:qifm}
\end{equation}
The
factor $R(\Delta_I)$ is the analogous of the clumping factor,
and is given by
\begin{equation}
R(\Delta_I) = \int_0^{\Delta_I} {\rm d} \Delta ~ 
\Delta^2 ~ P(\Delta)
\end{equation}
The reionization is complete when $Q_I = 1$; at this point a
mass fraction $F_M(\Delta_I)$ is ionized, while the rest is
(almost) completely neutral.

Note that this approach not only takes into account all the three stages 
of reionization, but also computes the clumping factor in a self-consistent
manner. This approach has been combined with 
the evolution of thermal and ionization properties of the IGM \cite{cf05}
to predict various properties related to reionization and then compare
with observations. By constraining the model free parameters with available
data on redshift evolution of Lyman-limit absorption systems,
GP and electron scattering optical depths, 
and cosmic star formation history, a unique
reionization model can be identified, whose main predictions are: 
Hydrogen was
completely reionized at $z \approx 15$, while helium
must have been doubly ionized by $z \approx 12$ by the metal-free
PopIII stars. Interestingly 
only 0.3 per cent of the stars produced by $z=2$ need to be PopIII stars in
order to achieve the first hydrogen reionization. 
At $z \approx 7-10$, the doubly ionized helium suffered an 
almost complete recombination as a result of
the extinction of PopIII stars.  A QSO-induced complete helium
reionization occurs at $z \approx 3.5$; a similar double hydrogen 
reionization does
not take place due to the large number of photons with energies $> 13.6$
eV from normal PopII stars and QSOs, even after all PopIII stars have
disappeared. Following reionization, the temperature of the
IGM corresponding to the mean gas density, $T_0$, is boosted to
$1.5 \times 10^4$~K; following that it decreases with a relatively flat
trend. 
Observations of $T_0$ are consistent with the 
fact that helium is singly ionized 
at $z \gtrsim 3.5$, while they are consistent with helium being 
doubly ionized at $z \lesssim 3.5$. 
This might be
interpreted as a signature of (second) helium reionization. 
However, 
it is useful to remember that 
there could be 
other contributions to the ionizing
background, like for example, because of 
thermal emission from gas shock heated during cosmic
structure formation \cite{mfwb04}. Such
emission is characterized by a hard spectrum extending well beyond 54.4 eV 
and is comparable to the QSO
intensity at $z \gtrsim 3$. These thermal photons alone could be enough
to produce and sustain double reionization of helium already at $z = 6$. 
The observations of the state of helium at these
intermediate redshifts ($3 < z < 6$) could be 
crucial to assess the nature of ionizing background arising from
different sources.

None of the above models take into account the clustering of the 
sources (galaxies) in computing the growth of the volume filling 
factor of ionized regions. 
It is well known that the galaxies would form in high-density regions 
which are highly correlated.
The overlap of the ionized bubbles and hence the morphology
of the ionized regions would 
then be determined by the galaxy clustering pattern; for example, 
the sizes of the ionized bubbles could be substantially underestimated
if the correlation between the sources are not taken into account. 
The first approach
to treat this has been to use an approach based on the 
excursion set formalism (similar to the Press-Schechter approach in spirit)
to calculate the growth and size distribution of ionized regions \cite{fzh04b}.
It can be shown that the sizes of the ionized regions can be larger than 
$\sim 10$ comoving Mpcs when the 
ionization fraction of the 
IGM is 0.5--1.0 \cite{fo05}. This highlights the fact that it is quite 
difficult to study overlap of ionized regions with limited box size in 
numerical simulations (which are often less that 10 comoving Mpc).
The other important result which can be drawn from such analyses is that
the reionization can be a very inhomogeneous process; 
the overlap of bubbles would be completed in different portions of the IGM at 
different epochs depending on the density inhomogeneities in that region.

\subsection{Simulations}

Though the analytical studies mentioned above allow us to develop a
good understanding of the different processes involved in reionization, they
can take into account the physical processes only in some approximate
sense. In fact, a detailed and complete description of reionization would
require locating the ionizing sources, resolving the inhomogeneities 
in the IGM, following the scattering processes through detailed radiative
transfer, and so on. Numerical simulations, in spite of their
limitations, have been of immense 
importance in these areas.

The greatest difficulty any simulation faces while computing the growth of 
ionized regions is to solve the radiative transfer equation \cite{anm99}
\begin{equation}
\frac{\partial I_{\nu}}{\partial t} 
+ \frac{{\bf n}  \cdot \nabla I_{\nu}}{\bar{a}} 
- H(t) \left(\nu \frac{\partial I_{\nu}}{\partial \nu} -3 I_{\nu}\right) 
= \eta_{\nu} - \chi_{\nu} I_{\nu}
\label{eq:rt}
\end{equation}
where $I_{\nu} \equiv I(t, {\bf x}, {\bf n}, \nu)$ is the monochromatic specific intensity of the radiation field, ${\bf n}$
is a unit vector along the direction of propagation of the ray, $H(t)$ 
is the Hubble parameter, and $\bar{a}$  is the ratio of cosmic scale factors between photon emission at frequency $\nu$ and the time $t$. Here $\eta_{\nu}$ and $\chi_{\nu}$ denote the emission coefficient and the absorption coefficient, respectively. The denominator in the second term accounts for the changes in path length along the ray due to cosmic expansion, and the third term accounts for cosmological redshift and dilution. 

In principle, one could solve equation (\ref{eq:rt}) directly for the intensity at every point in the seven-dimensional $(t,{\bf x}, {\bf n},\nu)$ space, given the coefficients $\eta$ and $\chi$. However, the high dimensionality of the problem makes the solution of the complete radiative transfer equation well beyond our capabilities, particularly since we do not have any obvious symmetries in the problem and often need high spatial and angular resolution in cosmological simulations. 
Hence, the approach to the problem has been to use different numerical schemes and approximations, like ray-tracing \cite{anm99,rs99,sah01,cen02,rnas02,sir04,bmw04,isr05,impmsa05},  Monte Carlo methods \cite{cfmr01,mfc03}, 
local depth approximation \cite{go97} and others.

Since the radiative transfer is computationally extremely demanding, most
efforts have been concentrating on small regions of space 
($\sim$ 10-50 comoving Mpc). 
The main reason for this limitation is that the ionizing photons 
during early stages of 
reionization mostly originate from smaller haloes which 
are far more numerous than the larger galaxies at high redshifts.
The need to resolve such small
structures imposes a severe limit on the computational box size. On
the other hand, these ionizing sources were strongly clustered at
high redshifts and, as a consequence, the ionized regions they created
are expected to overlap and grow to very large sizes, 
reaching upto tens of Mpc 
\cite{bl04,fo05,cen05}. As already discussed, the many orders of magnitude
difference between these length scales demand extremely high
computing power from any simulations designed to study early structure 
formation from the point of view of reionization. Further limitations are
imposed by the method used in the radiative transfer schemes; 
for most of them the computational expense grows roughly proportionally
to the number of ionizing sources present. This generally makes
the radiative transfer solution quite inefficient 
when more than a few thousand ionizing sources
are involved, severely limiting the computational volume that can
be simulated effectively.
However these methods can be useful and quite accurate in 
certain special
circumstances like, say, to study the growth
of ionized regions around an isolated source. 

A closely related problem which can be dealt with 
in numerical simulations is that
of the clumpiness of the IGM. We have already discussed on how the 
density inhomogeneities can play an important role in characterising 
reionization. Various hydrodynamical simulations 
have been carried out using sophisticated tools to generate the 
density distribution of gas over large ranges of spatial and density scales.
This, when combined with radiative transfer schemes, can give us an 
idea about the propagation of ionization fronts into an inhomogeneous medium, which is otherwise a very difficult problem.
Using radiative transfer simulations over large 
length scales ($\sim$ 150 comoving Mpc) based on 
explicit photon conservation in space and time \cite{impmsa05}, quite a few 
conclusions about the nature of reionization (by UV sources) can be drawn.
For example, it is likely that reionization proceeded in an inside-out 
fashion, with the high-density regions being ionized earlier, on average, than the voids. This has to do with the fact that most ionizing sources reside 
inside a high density halo. Interestingly, ionization histories of smaller-size (5 to 10 comoving Mpc) subregions exhibit a large scatter about the mean and do not describe the global reionization history well, thus showing the importance of large box sizes. The minimum reliable volume size for such predictions is found to be $\sim$ 30 Mpc. There seems to be two populations of ionized regions according to their size:  numerous, mid-sized ($\sim$ 10 Mpc) regions and a few, rare, very large regions tens of Mpc in size. The statistical distributions of the ionized fraction and ionized gas density at various scales show that both distributions are clearly non-Gaussian, indicating the non-linearities in the problem. 

There is thus quite good progress in studying the growth of the ionized
regions, particularly those due to UV sources, and also the qualitative
features for the global reionization seem to be understood well. The number of
unanswered questions still do remain large,  particularly those related
to the feedback effects. The expectation is that as more observations 
come up, the physics of the models would be nicely constrained and
a consistent picture of reionization is obtained. We shall review
the future prospects for this field in the next Section.

\section{The future} 

In the final Section, we review certain observations which will 
shape our understanding of reionization in near future, and also
discuss the theoretical predictions concerning future data sets.

The spectroscopic studies of QSOs at $z \gtrsim 6$ hold 
promising prospects for determining the neutrality of the IGM.
As we have discussed in Section 2, regions with high transmission in the Ly$\alpha$ forest 
become rare at high redshifts. Therefore 
the standard methods of analyzing the Ly$\alpha$ forest (like the 
probability distribution function and power spectrum) are 
not very effective. An alternative method to analyze the 
statistical properties of the transmitted flux is the distribution 
of dark gaps \cite{croft98,sc02}, defined as contiguous regions
of the spectrum having an optical depth above a threshold 
value (say 2.5 \cite{sc02} or 3.5 \cite{fsb++05}). 
The frequency and the width of the gaps 
are expected to increase 
with redshift, which is verified in different analyses of observational
data \cite{sc02,fsb++05}.
However, it is more interesting to check whether 
the dark gap width distribution (DGWD) is at all sensitive to the 
reionization history of the IGM, and whether one can constrain 
reionization through DGWD. This is indeed possible as it is found 
by semi-analytical models and simulations of Ly$\alpha$ forest
at $z \gtrsim 6$ \cite{gcf05}. 
In particular, about 30 per cent of the lines of sight (accounting for
statistical and systematic uncertainties) 
in the range $z= 5.7-6.3$ are expected to have 
dark gaps of widths 
larger than 60 \AA \ (in the QSO rest frame) if the IGM is in 
the pre-overlap stage 
at $z \gtrsim 6$, while no lines of sight should have such large gaps 
if the IGM is already ionized.
The constraints become more stringent at higher redshifts. 
Furthermore, 10 lines of sight should be sufficient for the DGWD to give statistically robust results and discriminate between early and late reionization scenarios. It is expected that the SDSS and Palomar-Quest survey \cite{ldc++05}
would detect $\sim 30$ QSOs at
these redshifts within the next few years and hence we expect 
robust conclusions from DGWD in very near future.

As we have discussed already, the first evidence for an early 
reionization epoch came from the CMB polarization data. This 
data is going to be much more precise in future with 
experiments like PLANCK,\footnote{http://www.rssd.esa.int/Planck/} 
and is expected to improve the 
constraints on $\tau_{\rm el}$. 
With improved statistical errors, 
it might be possible to distinguish between different evolutions
of the ionized fraction, particularly with E-mode polarization 
auto-correlation, as is found
from theoretical calculations \cite{hhkk03}.
An alternative option to probe reionization through CMB 
is through the small scale observations of temperature anisotropies.
It has been well known that the scattering of the CMB photons 
by the bulk motion of the electrons in clusters gives rise to a signal
at large $\ell$, known as the kinetic Sunyaev-Zeldovich (SZ) effect:
\begin{equation}
\Delta_T({\bf \hat{n}}) = \sigma_T \int {\rm d} \eta ~ 
{\rm e}^{-\tau_{\rm el}(\eta)} ~a
~ n_e(\eta,{\bf \hat{n}}) ~ 
\left[{\bf \hat{n} \cdot v}(\eta,{\bf \hat{n}})\right]
\end{equation}
where $\eta = \int {\rm d}t/a$ is the conformal time, 
${\bf v}$ is the peculiar velocity field and 
$n_e$ is the number density of electrons.
In principle, a signal should arise from the 
fluctuations in the distribution of free electrons 
arising from cosmic reionization. 
Now, if the reionization is uniform, the only fluctuations
in $n_e$ can arise from the baryonic density fluctuations $\Delta$, and
hence the power spectrum of temperature
anisotropies $C_{\ell}$ would be mostly determined by 
correlation terms like $\langle \Delta ~ {\bf v} ~ \Delta ~ {\bf v} \rangle$. 
Though this can give
considerable signal (an effect known as the Ostriker-Vishniac
effect), particularly for the non-linear densities, 
it turns out that for reionization the signal is dominated
by the patchiness in the $n_e$ distribution. In other words, 
if $\Delta_{x_e}$ denotes the fluctuations in the ionization
fraction of the IGM, the correlation term 
$\langle \Delta_{x_e} ~ {\bf v} ~ \Delta_{x_e} ~ {\bf v} \rangle$ 
(i.e., correlations of
the ionization fraction fluctuations and the large-scale bulk flow) 
gives the dominant contribution to the temperature anisotropies
$C_{\ell}$.
Now, in most scenarios of reionization, it is expected that 
the distribution of neutral hydrogen would be quite 
patchy in the pre-overlap era, with the ionized hydrogen 
mostly contained within isolated bubbles. 
The amplitude of this signal is significant around 
$\ell \sim 1000$ and is usually comparable to or
greater than the signal arising from standard kinetic SZ effect 
(which, as mentioned earlier, is related to the scattering of the CMB photons 
by the bulk motion of the electrons in clusters). 
Theoretical estimates of the signal have been performed 
for various reionization scenarios, and it has been predicted that
the experiment can be used for constraining reionization history
\cite{schkm03,mfhzz05}. 
Also, it is possible to have an idea about the nature of reionization
sources, as the signal from UV sources, X-ray sources and 
decaying particles are quite different. With multi-frequency 
experiments like 
Atacama Cosmology Telescope (ACT)\footnote{http://www.hep.upenn.edu/act/} 
and South Pole Telescope (SPT)\footnote{http://spt.uchicago.edu/}
coming up in near future, this promises to 
put strong constraints on the reionization scenarios.

Another interesting prospect for constraining reionization is
through high redshift energetic sources like gamma ray bursts (GRBs)
and supernovae. There are different ways of using these sources
for studying reionization.
The first is to study the spectra of individual sources and 
estimate the neutral fraction of hydrogen through its
damping wing effects. This is similar to what is done in the 
case of Ly$\alpha$ emitters as discussed in Section 2. 
The damping wing of the surrounding neutral medium, 
if strong enough, would suppress the spectrum 
at wavelengths redward of the Lyman break. In fact, analyses
have been already performed on the GRB with highest detected
redshift ($z_{\rm em} = 6.3$), and the wing shape is well-fit
by a neutral fraction $x_{\rm HI}  = 0.00 \pm 0.17$ \cite{tkk++05}. 
In order to obtain more stringent limits on reionization, 
it is important to increase the sample size of $z >6$ GRBs.
Given a reionization model, one can actually calculate the
number of GRB afterglows in the pre-reionization era 
which would be highly absorbed
by the neutral hydrogen. 
These GRBs would then be categorised as
``dark'' GRBs (i.e., GRBs without afterglows), and the 
redshift distribution of such objects can give us a good handle
on the evolution of the neutral hydrogen in the universe \cite{cs02,bl02}.

The second way in which GRBs could be used is to constrain the 
star formation history, and hence get indirect constraints on  
reionization. In most popular models of GRBs, it is assumed that they
are related to collapse of massive stars (just like supernovae), 
and hence could be 
nice tracers of star formation. 
In fact, one can write the number of GRBs (or supernovae) per 
unit redshift range observed over a time $\Delta t_{\rm obs}$ as
\cite{cs02,mjh06}
\begin{equation}
\frac{{\rm d}  N}{{\rm d}  z} = \frac{{\rm d}  \Omega}{4 \pi}
\Delta t_{\rm obs} \Phi(z) ~ \left[f \dot{\rho}_{\rm SF}(z)\right] 
\frac{1}{1+z} \frac{{\rm d}  V(z)}{{\rm d}  z}
\end{equation}
where the factor $(1+z)$ is due to the time dilation between $z$ and the 
present epoch, ${\rm d}  V(z)$ is the comoving volume element, 
${\rm d}  \Omega/4 \pi$ is the mean 
beaming factor and $\Phi(z)$ is the 
weight factor due to the limited sensitivity of the detector, because of 
which, only brightest bursts will be observed at higher redshifts.
The quantity
$f$ is an efficiency factor which links the formation of
stars $\dot{\rho}_{\rm SF}(z)$
to that of GRBs (or supernovae); it corresponds to the number of GRBs (supernovae) 
per unit mass of forming stars, hence it depends on the fraction of 
mass contained in (high mass) stars which are potential progenitors 
of the GRBs (supernovae). Clearly, the value of $f$ might depend
on some details of GRB formation and is expected to be quite
sensitive to the stellar IMF.
Though such details still
need to be worked out, it seems promising that data on the 
redshift distribution of GRBs and supernovae could give a handle
on the star formation rate, which in turn could give us insights
on quantities like efficiency of molecular cooling or the 
relative contribution of minihaloes to radiation.
In general, the GRB rates at high redshifts should be able 
to tell us how efficient stars were in ionizing
the IGM.

Perhaps the most promising 
prospect of detecting the fluctuations in the neutral 
hydrogen density during the (pre-)reionization era is through 
the future 21 cm emission
experiments \cite{cm03} like LOFAR\footnote{http://www.lofar.org}. 
The basic principle which is central to these experiments is the 
neutral hydrogen hyperfine transition line at a rest wavelength of 
21 cm. This line, when redshifted, is observable
in radio frequencies ($\sim 150$ MHz for $z \sim 10$)
as a brightness temperature:
\begin{equation}
\delta T_b(z, {\bf \hat{n}}) = \frac{T_S - T_{\rm CMB}}{1+z} 
~ \frac{3 c^3 \bar A_{10} n_{\rm HI}(z,{\bf \hat{n}}) (1+z)^3}
{16 k_{\rm boltz} \nu_0^2 T_S H(z)}
\end{equation}
where $T_S$ is the spin temperature of the gas, $T_{\rm CMB} = 2.76 (1 + z)~ {\rm K}$ is the CMB temperature, $A_{10}$ is the Einstein coefficient
and $\nu_0 = 1420$ MHz is the rest frequency of the hyperfine line.
The expression can be simplified to
\begin{equation}
\delta T_b(z, {\bf \hat{n}}) = F_0 ~ 
\frac{T_S - T_{\rm CMB}}{T_S}
~ \frac{n_{\rm HI}\left(z,{\bf \hat{n}}\right)}
{\bar{n}_{\rm H}(z)}
\label{eq:t_b}
\end{equation}
with
\begin{equation}
F_0 =  \frac{3 c^3 \bar A_{10} \bar{n}_{\rm H}(z) (1+z)^2}
{16 k_{\rm boltz} \nu_0^2 H(z)}
\approx 25 {\rm mK} \sqrt{\frac{0.15}{\Omega_m h^2}}
~\left(\frac{\Omega_B h^2}{0.022}\right)
\left(\frac{1 - Y}{0.76}\right) \sqrt{\frac{1+z}{10}}
\end{equation}

The observability of this brightness temperature against the 
CMB background will depend on the relative values 
of $T_S$ and $T_{\rm CMB}$. Depending on which processes
dominate at different epochs, $T_S$ will couple 
either to radiation ($T_{\rm CMB}$) or to matter (determined
by the kinetic temperature $T_k$). There are four broad 
eras characterising the spin temperature
\cite{wouthuysen52,field59}: (i) At $z \gtrsim 30$, the 
density of matter is high enough to make collisional
coupling dominant, hence $T_S$ is coupled to 
$T_k$. However, at $z \gtrsim 100$, the gas
temperature is coupled strongly to $T_{\rm CMB}$, thus
making $T_S \simeq T_k \simeq T_{\rm CMB}$. At these epochs, the 
21 cm radiation is not observable. (ii) At $30 \lesssim z \lesssim 100$, 
the kinetic temperature falls off adiabatically and hence is lower than
$T_{\rm CMB}$, while $T_S$ is still collisionally coupled
to $T_k$. This would imply that the 21 cm radiation will 
be observed in absorption against CMB. (iii) Subsequently
the radiative coupling would take over and make $T_S = T_{\rm CMB}$,
thus making the brightness temperature vanish. This continues
till the sources turn up and a Ly$\alpha$ background is established.
(iv) Once there is background of Ly$\alpha$ photons, that 
will couple $T_S$ again to $T_k$ through the Wouthuysen-Field mechanism. From 
this point on, the 21 cm radiation will be observed either
in emission or in absorption depending on whether 
$T_k$ is higher or lower than $T_{\rm CMB}$, which turns out
be highly model-dependent.

Almost in all models of reionization, the most interesting phase
for observing the 21 cm radiation is $6 \lesssim z \lesssim 20$. This 
is the phase where the IGM is suitably heated to temperatures
much higher than CMB (mostly due to X-ray heating \cite{cm04}) thus making
it observable in emission. Furthermore, this is the era when the 
bubble-overlapping phase is most active, and there is substantial
neutral hydrogen to generate a strong enough signal. At low
redshifts, after the IGM is reionized, $n_{\rm HI}$ falls by orders
of magnitude and the 21 cm signal vanishes.

Most theoretical studies are concerned with studying the angular
power spectrum of the brightness temperature fluctuations, which
is essentially determined by the correlation terms 
$\langle \delta T_b~  \delta T_b \rangle$. It is clear from equation
(\ref{eq:t_b}) that the temperature power spectrum is 
directly related to the power spectrum of neutral hydrogen, i.e., 
$\langle n_{\rm HI} ~ n_{\rm HI} \rangle$. This then turns 
out to be a direct probe of the neutral hydrogen distribution, and
potentially can track the evolution of the patchiness in the distribution
over redshift. 
In fact, one expects a peak in the signal on angular scales 
corresponding to the characteristic size of the ionized bubbles.
While there are some significant systematics which have
to be controlled (say, for example, the foregrounds), the
experiments do promise a revolution in our understanding of reionization.

There are interesting ways in which one can combine signals from
different experiments too. For example, an obvious step would be 
to calculate the correlation between CMB signal from 
kinetic SZ effect and the 21 cm brightness temperatures
$\langle \Delta_T ~ \delta T_b \rangle$. This will essentially
be determined by the correlations 
$\langle n_e ~ n_{\rm HI} ~ {\bf v} \rangle$ \cite{cooray04}. It is 
expected, that the ionized number density $n_e$ will 
be highly anti-correlated with  the neutral 
number density $n_{\rm HI}$. In fact, the simulations do 
show a clear signal for this anti-correlation. Depending on the 
angular scales of anti-correlation, one can actually re-construct the 
sizes of the bubbles as a function of redshift and thus 
compute the reionization history \cite{scfb05}. 

We hope to have convinced the reader that we are about 
to enter the most exciting phase in the study of reionization as 
new observations 
with LOFAR, ALMA and NGWST will soon settle the long-standing question
on when and how the Universe was reionized. From the theoretical point of view, 
it is thereby important to
develop
detailed analytical and numerical models to extract the maximum
information about the physical processes relevant for reionization 
out of the expected large and complex data sets.

\small{
 
}

\end{document}